# A Systematic Review of Machine Learning Enabled Phishing

KRYSTAL A JACKSON

## Abstract

Developments in artificial intelligence (AI) are likely to affect social engineering and change cyber defense operations. The broad and sweeping nature of AI's impact means that many aspects of social engineering could be automated, potentially giving adversaries an advantage. In this review, we assess the ways phishing and spear-phishing might be affected by machine learning techniques. By performing a systematic review of demonstrated ML-enabled phishing campaigns, we take a broad survey the space for current developments. We develop a detailed approach for evaluation by creating a risk framework for analyzing and contextualizing these developments.

The object of this review is to answer the research questions:

(1) Are there high-risk ML-enabled phishing use cases?
(2) Is there a meaningful difference between traditional targeted phishing campaigns and ML-enabled phishing campaigns?

Practitioners may use this review to inform standards, future research directions, and cyber defense strategies.

# Introduction and Background

Defenders are always looking to prepare for new threats, particularly those posed by emerging technologies. Machine learning has proven to be an essential asset in cyber defense operations for decades, from spam filtering to intrusion detection systems [1]. Slightly less explored has been the effect of machine learning on offensive operations and social engineering. Some researchers have theorized about the effect of machine learning on the cyber threat landscape. Theories include next-generation "smart malware" capable of dynamically learning a victim's computing environment and mimicking trusted user behavior [2][3], systems that can thwart bot detection such as CAPTCHA [4], and systems that can drastically improve phishing attacks [5]. Researchers have already demonstrated that some of these systems can be designed and deployed. Phishing in particular is a growing threat. Among cases reported to the FBI, phishing affected over 300,000 victims resulting in tens of millions in losses in 2021 [6]. However, for each of these demonstrations, it is essential to consider the associated risk and potential impacts. The security research community has different capabilities, intent, and access than adversaries. Taking these demonstrations at face value, we extract relevant information and contextualize potential impacts. This review presents the security research community with an overview of ML-enabled phishing and spear-phishing developments. The goal is to place demonstrated and theorized attacks and capabilities into context with a clear picture of their risks and consequences.

**This paper is organized into 8 sections:**
Section 1 <u>Definitions</u>: Established the terminology used throughout this paper.
Section 2 <u>Anatomy of a Phish</u>: Provides a taxonomy for understanding modern phishing campaigns. This is not comprehensive but allows us to understand how current phishing operations are conducted and where machine learning may impact different aspects of those operations.
Section 3 <u>Research Questions and Objectives</u>: Describes our two research questions and their rationale.
Section 4 <u>Methodology</u>: Described the review process including how studies were identified and filtered.
Section 5 <u>Case Study Analysis:</u> Describes how each case study was evaluated and the risk analysis process used.
Section 6 & 7 <u>Discussion of Research Questions</u> and <u>Synthesis Discussion:</u> Presents the findings and answers our research questions.
Section 8 <u>Future Work:</u> Discusses avenues for future research.

# Definitions

Phishing

**Phishing**

Lastdrager, in their review of phishing definitions, found that a definitional consensus could be reached by defining phishing as, 'a scalable act of deception whereby impersonation is used to obtain information from a target' [7]. Although general enough to be widely agreed on, we prefer the more specific definition by Singh and Somase who define phishing as "Phishing is the criminally fraudulent process of attempting to acquire sensitive information such as usernames, passwords, and credit card details, by masquerading as a trustworthy entity in an electronic communication"[7]. We modify this definition slightly to, "Phishing is a criminally fraudulent process whereby malicious actors masquerading as a trustworthy entity use text-based methods of deception to acquire sensitive information."

**Spear-Phishing**

We define spear-phishing as, "Spear-phishing is a criminally fraudulent process whereby malicious actors masquerading as a trustworthy entity use targeted text-based methods of deception to acquire sensitive information." Spear-phishing can use open-source intelligence gathering (OSINT) or other methods of reconnaissance. This information provides the attacker with enough context to make the scam portion of their attack appealing enough such that the target will fall for their phishing attempt.

While a phishing campaign aims to be general, with the goal that some percentage of those targeted are successfully scammed, spear-phishing is aimed at scamming one individual or a very targeted subset of individuals. The information adversaries are after with either type of phishing attack is typically sensitive or private information. The information may or may not be the end goal in and of itself. Bank login details, for example, might be used by an adversary to access accounts for the purpose of initiating a wire transfer. Adversaries typically want to perform some action with the information they obtain such as gaining unauthorized access, making sensitive information public, or using it for further deception.

**Other Distinctions**

Another popular term, 'whaling', is meant to describe a highly targeted attempt to gather information from important individuals. For the purposes of this research, we see no meaningful difference between highly targeted spear-phishing and whaling and therefore stick to the phishing and spear-phishing distinction.

The distinction between spam and phishing is also of interest. In their book, *Spam: A Shadow History of the Internet*, Brunton defines spam as, "spamming is the project of leveraging information technology to exploit existing gatherings of attention" [8]. Typically we consider spam to be bulk messages sent by commercial entities. The goal of spam might be to get one to take an action, such as making a purchase, but it is less likely to be extractive in nature. This is the key difference between spam and phishing and why phishing is known for adversaries "fishing for information." A spam message may hide the sender's identity but is more likely to be promoting or highlighting it instead. Furthermore, while spam may use deception and manipulation, such as forwarded hoax email chains for spreading misinformation, they are less extraction-oriented.

**Traditional Phishing and ML-enabled Phishing**
ML-enabled phishing is any phishing operation that involves the automation or execution of tasks using machine learning tools and techniques. This might look like creating what we call the *scam* portion of the phish by using natural language processing (NLP). It might also look like learning what email headers bypass filtering or what group of users on a social media platform are most susceptible to a phish by using long short-term memory networks. What a traditional phishing campaign looks like and more details on where ML may influence the phishing lifecycle are presented in the *Anatomy of a Phish* and *Synthesis Discussion* sections.

As task automation becomes more commonplace, the distinction between ML-enabled phishing and traditional phishing may become less meaningful. Any technology as ubiquitous as artificial intelligence will likely influence phishing and other social engineering operations as it has influenced all of cybersecurity. Ultimately, while we make this distinction here, what we attempt to offer is a nuanced discussion of what influence machine learning will have on current cyber defense operations. However, machine learning is an ever-changing technology. Advances in chip design, algorithms, and increasingly lower data storage costs, could lead to a new set of tools and techniques defenders will need to prepare for. One of the most important things security researchers can do is focus on understanding machine learning capabilities. This aids in the establishment of standards with which the risk of new developments can be assessed. In addition to analyzing each case study of ML-enabled phishing, we also provide a risk analysis framework in the *Case Study Analysis* and *Synthesis Discussion* sections.

Risk

**Risk: Threat X Vulnerability X Consequences**
Formally, a *risk* is a measure of a *threat* exploiting a *vulnerability* to cause some *impact* on a victim or organization (victims). The vulnerability can be a technical one, such as a lack of an email filter, or a policy or managerial one, such as a lack of email security training. A vulnerability existing by itself is not enough and must be accompanied by a threat willing to leverage it to be actionable. By leveraging this vulnerability, the adversary has some impact on an individual or organization measured by the consequences the victim faces.

**Threat: Intent X Capability X Access**

When we discuss a threat in this work we are describing an outside malicious threat actor. Other threats can exist but are outside the scope of this research. A threat actor can be broken down further by their capability, intent, and access. The capability of a threat actor is measured by their available resources and knowledge. Access is the opportunity or ability of the threat actor to reach their target. For example, even with the resources and knowledge to conduct an ML-enabled phishing campaign on Twitter, if a user is simply not on the platform the scam will never reach them. The intent is measured by a threat actor's desire or motivations. Our research assumes intent, meaning that we assume there are adversaries with the intention of conducting phishing attacks and with an interest in doing so as efficiently as possible. Therefore, if machine learning gave a competitive advantage in phishing campaigns we assume adversaries would be interested in using it. Our main focus then is on analyzing if this is indeed the case, and if so, what capabilities, resources, and access the threat actors would need.

# Anatomy of a Phish

To better understand the influence of machine learning on phishing we outline the basic anatomy of phishing messages as they are carried out over various attack vectors. A phishing attack is composed of three parts, the **lure**, the **scam**, and the **delivery**. Depending on the objectives and resources of the adversary, these three categories can look very different between different attacks. A **lure** is any part of the phishing campaign that grabs the attention of the victim. A **scam** is typically the subject line and body of the phishing attempt. The **Delivery** is the link, malicious document, or other executable that the adversary wants the victim to interact with.

We have separated the lure from the scam to make an important distinction between the elements that initially present themselves to a victim and the elements that motivate action on the part of the victim. Not all phishing scams involve a delivery, for example, an adversary might motivate a phone call or wire transfer as part of a business email compromise instead. This outline is in no way meant to be comprehensive or cover all aspects of a phishing campaign. We have omitted the reconnaissance position of the phishing lifecycle in this description, but address some questions related to information gathering in the *Case Study Analysis* and *Synthesis Discussion* sections. Information gathering is an essential part of the phishing lifecycle, and can involve gathering email lists, information on targets for personalized attacks, or searching for vulnerabilities within an organization. ML could aid in all of these processes, however, for the scope of this research, we have chosen to focus on only a few aspects of reconnaissance that came up in our case studies.

### Lure

The lure is the mechanism or vehicle employed during the phishing campaign. For example, spoofed email headers made to look like communications from a trusted source. Email spoofing is when the information in the Mail From, From, Reply-To, or Date section of the email is falsified, making it appear as if it is coming from or should go to a legitimate source [9]. Lures are typically either in an email, direct message over social media, or SMS (also known specifically as SMiShing) [10]. In phishing defense training, people are typically instructed to inspect the lure first for signs of fabrication. However, a more thorough investigation of any communications is usually triggered by suspicion from something in the scam portion [11].

Email headers are made up of information about the sender, the 'From:' address, a timestamp, IP address, as well as information from the various message transfer agents (MTA)s involved in the delivery of the email [12]. Most users will only see the 'From:' field which can easily be spoofed and the timestamp.

**The Machine Learning Impact**
It is theorized machine learning could impact the lure portion of a phishing campaign if adversarial ML models are used to help trick current filtering techniques [13].

### Scam

Or hook; this is the information that motivates a victim to take action. Scams are contained in the subject line and body of an email, or make up the text of an SMS or direct message. Scams rely on a few psychological principles to be successful, namely creating a feeling of immediacy, communicating failure, using an authoritative tone, and expressing a shared interest [14] [15]. The approaches used to scam, manipulate, and deceive someone, have stayed relatively consistent over human history and mediums [14]. The literature most often referenced when discussing the psychological principles at play with phishing scams include "Cialdini's principles of influence, Gragg's psychological triggers, and Stajano *et al.* principles of scams" [14]. These taxonomies overlap in many ways and we will focus on their commonalities and applications to social engineering. In social engineering, the authors of *Principles of Persuasion in Social Engineering and Their Use in Phishing* found that the principles of 'Liking, Similarity & Deception' (LSD) are used across most types of phishing emails [14]. The authors explain that these principles are fundamental to the ways humans interact socially by tapping into our desire to try and "connect with others by finding characteristics that are more agreeable and similar to them. Humans tend to believe in what other humans do or say unless they suspect something is really wrong or that some behavior is completely unexpected" [14].

**The Machine Learning Impact**
With the increase in ability from NLP comes the risk of computer-generated text that can fool humans. The question, can a computer pass the Turing test or simply fool or impersonate a human, has yet to be answered. Given the current state of the art NLP, it is possible to generate *human-like* text, with a seed or prompt [16]. This text is far from infallible though and often requires editing or modification [16]. For an NLP to be meaningfully said to be able to fool another person requires an understanding of human psychology that far surpasses even the most advanced tools available to us today. For the time being, NLP is a powerful text generation tool that, like many other tools, is only as smart as the user wielding it. NLP does have the capacity to improve the efficiency at which deceptive messages can be generated, but is highly case dependent [17].

The other area of concern stems from ML tools used to psychologically profile targets. Tools like Crystal Knows and Humantic AI which are used by recruiters can be used to assess the psychological disposition of a candidate for outreach and can be used to profile victims for phishing attacks [18]. These tools could be used by adversaries to better inform the tone of a scam to be more receptive to their target. These tools only provide an advantage in the case of target phishing campaigns, since each message would theoretically need to be personalized in this way.

## Delivery

This is where the victim gets "reeled" in. Depending on the attack, the delivery could be things such as a URL that directs the victim to a page to reveal sensitive information or a malicious attachment that infects a system. We typically do not think of 'calls to action' as deliveries. An email instructing a victim to place a phone call or reply to an email with information is not delivering anything. Not all phishing attacks contain a delivery which makes assessing the relationship between attacks difficult. With deliveries that rely on the victim simply opening an attachment, an attacker may not even care about the scam portion of the phish, spending time only on the lure, and in particular the subject line, to make it just enticing enough for a victim to investigate the attachment [19].

**The Machine Learning Impact**
ML might be capable of learning what content is filtered or dynamically learn which URLs perform the best in campaigns, as demonstrated in one of our case studies [20].

# Research Questions and Objectives

| Research Questions | Rationale |
| --- | --- |
| R1: Are there high-risk ML-enabled phishing use cases? | If ML-enabled phishing carries risks, what is the impact and likelihood of those risks being realized? Are there high-impact risks defenders can prepare for, and if so, in what ways. |
| R2: Is there a meaningful difference between traditional phishing campaigns and ML-enabled phishing campaigns? | What difference, if any, do ML-enabled phishing campaigns have. If the techniques and methods differ significantly, in what ways. |

**Table 1** *Research Questions*

To answer these questions, we conduct a review of existing literature, evaluate our findings using a risk framework, and synthesize the results across findings. Systematic reviews allow practitioners to keep apprised of developments in their field. In addition, these reviews often function as a starting place for developing evidence-driven guidelines and norms. By synthesizing information across multiple studies with different designs, we can begin to get a larger picture of this emerging concern.

# Methodology

This study's methodology was informed by *How to Do a Systematic Review: A Best Practice Guide for Conducting and Reporting Narrative Reviews, Meta-Analyses, and Meta-Syntheses* by Siddaway et al [21]. Although intended for a psychology research audience, this methodology provided a comprehensive method for conducting a narrative review as opposed to a meta-review. In addition, we utilized the *PRISMA guidelines for reporting systematic reviews* for guidelines on reporting the search procedure [22].

The review process includes evaluating each finding (henceforth called a case study) against a series of binary questions. To synthesize our findings, we adapted a risk framework from the *National and Transnational Security Implications of Big Data in the Life Sciences* from the joint AAAS-FBI-UNICRI project [23]. The primary focus is to assess these studies to determine capabilities and access, which will provide a measure of threat. Synthesis allows us to get a larger picture of ML-enabled phishing and in the *Synthesis Discussion* section, we will explore some of the main takeaways from this analysis.

## Scope

Our scope included English language publications from five different information security conferences, Black Hat, FIRST, Usenix, IEEE S&P, Def Con, and a search of Semantic Scholar and Google Scholar. We limited our search to include the newest case studies only from 2018-2022. Both demonstrated attacks with empirical results and theoretical attacks were evaluated. We searched for all keyword terms on each platform shown in **Table 2**, and we created this list by using synonyms and related words to our focus "machine learning, adversarial, phishing."

Terms used for search across publications and databases

| |
|---|
| malicious ML, malicious ML for phishing, adversarial ML, adversarial ML phishing, automated phishing, AIaaS phishing, phishing using ML, phishing using AI, ML phishing attacks, AI phishing attacks, offensive AI phishing, offensive ML phishing, advances in phishing techniques, ML phishing techniques, automated spear-phishing, advances in ML and phishing, weaponized machine learning phishing, machine learning-enabled phishing, information operations |

***Table 2** Search Terms*

## Data Extraction Process

From each case study we extracted information to inform a risk scoring for that study shown in **Table 3**. Each question was recorded as a binary 'yes/no' response and the sum of 'yes' responses totaled for each case study. Unknown information is treated the same as a 'no' for the risk scoring. From there, each study was marked as high, medium, or low risk, according to the risk matrix in **Figure 1**.

### Inclusion Criteria

Studies within the scope that were included also met the following criteria:
   (1) A keyword from Figure 1 was included in the title or abstract
   (2) The case study focused on using machine learning methods for enhancing, enabling, or otherwise carrying out the phishing attack
         (a) Machine learning including the subfields of deep learning and neural networks

### Exclusion Criteria

Studies within the scope that were excluded met the following criteria:
   (1) Case study used machine learning methods for improving phishing detection as opposed to attacks
   (2) Same case study from a different source

(3) Case study was relevant to search criteria but was brief enough to not address research questions or allow us to perform extraction on

## Case Studies

We identified 4 case studies for further evaluation. These case studies were:
  (1) *Turing in a Box: Applying Artificial Intelligence as a Service to Targeted Phishing and Defending against AI-generated Attacks* [18]
  (2) *Generative Models for Spear Phishing Posts on Social Media* [24]
  (3) *OFFENSIVE AI: UNIFICATION OF EMAIL GENERATION THROUGH GPT-2 MODEL WITH A GAME-THEORETIC APPROACH FOR SPEAR-PHISHING ATTACKS* [28]
  (4) *DeepPhish: Simulating Malicious AI* [20]

Each case study had a different level of detail and varied widely in the quality and depth of information. Nonetheless, we evaluated each against the same criteria and took note of when information was not clear or missing from a case study.

## Case Study Analysis

Using the extracted information from the case studies, we assigned each a risk score. There are 5 questions regarding impact with 7 measures of impact and 7 measures of likelihood across 12 total questions. In our assessment, these questions are the minimum required to gain an understanding of the risk of ML-enabled phishing. One could argue for or against any of these questions and several have aspects of both impact and likelihood to consider. However, since all questions go into calculating the final risk scoring we believe the distinction is not of paramount importance to the final analysis. We encourage further research to build upon and modify these risk questions.

| Impact | Likelihood |
| --- | --- |
| Does this study produce results that are more effective than traditional phishing campaigns? | Are known defensive strategies against traditional phishing ineffective? |
| How was the attack carried out?<br>- Email<br>- SMS<br>- Social Media | Are any new defensive strategies mitigating the ML-enabled attack presented in the case study? |

| Scam Type?<br>- LSD | Was the delivery generated in a typical way? |
|---|---|
| Was there a Delivery? | Publicly available model(s) used? |
| Were models used for OSINT? | Are publicly available data set(s) used? |
|  | Did targeting happen without specialized access to targets? |
|  | Did it forgo specialized methods for grouping targets? |

*Table 3* Extraction Questions

Each case study's overall risk was assessed using the following matrix. The cuts off for the low and high-risk scoring are included.

| Risk Likelihood | Risk Impact | |
|---|---|---|
|  | HIGH: >= 4 | LOW: <4 |
| HIGH: >= 4 | HIGH | MEDIUM |
| LOW: < 4 | MEDIUM | LOW |

*Table 4* Case Study Risk Matrix

## Impact

**Does this study produce results that are more effective than traditional phishing campaigns?**
It is estimated that phishing campaigns have a success rate, measured as a 'click-through' rate, of approximately 7-14% on average [24]. Click-through rates measure a victim opening and engaging with the phish in some way. Typically this measure is used for phishing emails that contain a malicious document or URL the victim must interact with. It could also measure an email response or phone call back to an adversary providing the requested information. Since this term is limited in some respects and less intuitive for some cases of phishing it is difficult to report the exact rate of phishing success. For example, an 'open-rate' may not seem as meaningful a measure if a victim does not provide the information the adversary was after, unless that message contained some type of malicious file that was executed upon opening. However, for the purposes of our research, we focus less on the distinction between open and click-through rates. This is because most case studies in our analysis claimed to have success rates far above the average rates for either open rates or click-through rates, and marginal improvements would not cause ML-enabled phishing to rise to the level of a high-risk use case or present a meaningful difference between traditional and ML-enabled phishing.

**How was the attack carried out?**
For each possibility, email, SMS, or social media, we recorded a binary 'yes/no' response. If an attack can be carried out over multiple attack vectors we assign it a higher risk score.

**Scam Type?**
As discussed in the *Autonomy of a Phish: Scam* section, the psychological manipulation tactics used impact whether a victim falls prey to an attack or not [14]. As best as can be determined (and when relevant) we extract the methods used in these studies to see if they rely on the principles of Liking, Similarity & Deception (LSD). However, most studies do not go into detail about their scam tactics. Since this is a critical part of assessing the impact of NLP in particular on the effectiveness of phishing campaigns we identify this as a key area of further research.

**Was there a Delivery?**
Without a Delivery portion of a phish, it is impossible to measure click-through rates. Additionally, without this portion, a phish may be more likely to make it through any email filtering techniques.

**Were models used for OSINT?**
In the process of creating a phishing campaign, particularly in spear-phishing campaigns, adversaries will gather information on their target(s) to do things such as personalize their messages or time the delivery of their attacks [15]. With this question, we extract information about the use of ML for automating part or all of the intelligence-gathering process. We assume highly targeted messages will result in higher success rates as opposed to mass phishing [14][26].

## Likelihood

**Are known defensive strategies against traditional phishing effective to any degree?**
In some of the case studies, it is noted that defense strategies such as email filtering were effective against the ML-enabled attacks [18] [20]. If known defense strategies are effective it will be important to get a clearer picture of; to what degree, in what ways, and in which cases.

**Are any new defensive strategies mitigating the ML-enabled attack presented in the case study?**
It is important to explore what new developments are needed to meet the challenges of detecting ML-enabled attacks. If significant new developments are not needed, the likelihood of a successful attack and overall risk decreases. In *Automatic Detection of Machine Generated Text: A Critical Survey*, for example, it is theorized that there are several methods of identifying machine-generated text [25].

**Was the delivery generated in a typical way?**
If a delivery was present, was it created in the same way a delivery in a traditional phishing campaign would have been?

**Publicly available model(s) used?**
In the *Synthesis Discussion* section we elaborate on the factors motivating adversaries and explain why we assume they operate as an instrumentally rational actors. We consider a publicly available model as any model developed by an entity that is easily accessible by the general public for use. If an adversary can quickly and efficiently use ML models that are publicly available to conduct their campaigns then the cost to them is lessened. However, if models must be built and trained, this increases the cost to adversaries significantly, and we can assume the average adversary would be unwilling or unable to do such a process.

**Are publicly available data set(s) used?**
Similarly to the question of publicly available models, we ask if publicly available data sets are used. We consider publically available data as any data created or gathered by an entity that is easily accessible by the general public for use. Some models require more data than others for training and testing.

**Did targeting happen without specialized access to targets?**
Examples of specialized access include permissions to follow an account on social media or the ability to send emails from a pre-approved or white-listed account. Researchers with specialized access may have an advantage over adversaries.

**Did it forgo specialized methods for grouping targets?**
One possible use of ML for phishing campaigns is target selection. Potential targets can be grouped based on a variety of behavioral factors. ML could also be used to group targets if enough data were available to discern a meaningful difference between targets. Although some literature theorizes about what factors make an individual susceptible to phishing, there is still little consensus on this topic and any other variables involved [26]. Grouping targets requires additional resources and does not have strong empirical evidence to support that this practice greatly increases phishing access rate given the variety of variables involved in selecting and sorting targets.

The results of this analysis of each study are captured below

| Case Study | Impact | Risk Impact | Likelihood | Risk Likelihood | Overall Risk Scoring |
|---|---|---|---|---|---|
| Turing in a Box: Applying Artificial Intelligence as a Service to Targeted Phishing and Defending against AI-generated Attacks | 3 | LOW | 2 | LOW | LOW |
| Generative Models for Spear Phishing Posts on Social Media | 4 | HIGH | 4 | HIGH | HIGH |
| OFFENSIVE AI: UNIFICATION OF EMAIL GENERATION THROUGH GPT-2 MODEL WITH A GAME-THEORETIC APPROACH FOR SPEAR-PHISHING ATTACKS | 2 | LOW | 1 | LOW | LOW |
| DeepPhish: Simulating Malicious AI | 4 | HIGH | 3 | LOW | MEDIUM |

*Table 5* Results of Case Study Analysis

# Discussion of Research Questions

## R1

**Are there high-risk ML-enabled phishing use cases?**
From our analysis we conclude that there is **one case** of high-risk ML-enabled phishing. The combination of triaging targets, conducting an attack on social media, in particular Twitter where 'textspeak' is commonplace, shortened URLs, and use of open models, make this a high-risk case of ML-enabled phishing.

The authors leverage multiple tactics to improve their phishing campaign. This case is best analyzed as an attempt to understand how modern communication tools and techniques can aid adversaries. Seymour and Tully use two different machine learning models, one for triaging users and one for generating text and a simulated phishing URL. They also employ several in-depth targeting methods, selecting users who have a high phishing susceptibility. They measure this through metrics such as high follower count, high post count, descriptive bio text including job descriptions and company names, recent profile creation dates, and many changes from the default account settings [24]. Although a seemingly reasonable tactic, they do not provide justification for why these factors increase a target's phishing susceptibility. Whether these factors are indicative of phishing susceptibility on Twitter or other social media platforms is a question we believe requires further research. Seymour and Tully also collect user post history to generate the targeted message. They pre-train a long short-term memory (LSTM) model using the Twitter GloVe embeddings. Additionally, there is mention of attempting several methods for extracting a seed for the generated text from a user's profile, but not much information is provided on this part of the process. It is reasonable to assume that, given their discussion of their model and natural language processing, they attempt to use NLP to identify topics each target might be interested in and use some string of information from user profiles as a seed.

Furthermore, their work is an example of combining the targeted spear-phishing approach with the en masse approach of traditional phishing campaigns. It is reasonable to believe as the tools and techniques used to target individuals become more accessible, more adversaries will utilize them. However, the risk of large-scale psychological targeting based on social media data has already been realized multiple times. Matz et al. define psychological targeting as "the practice of influencing the behavior of large groups of people through interventions that are psychologically tailored" [27]. In 2016, for example, Cambridge Analytica used data from approximately 230 million Facebook (now Meta) profiles to influence the 2019 Presidential election [28]. The combination of big data and the plethora of tools for scraping, aggregating, and sorting personal information is a growing challenge for the security community that should not be ignored.

## R2

**Is there a meaningful difference between traditional phishing campaigns and ML-enabled phishing campaigns?**
From our analysis we conclude that there is **NOT** a meaningful difference between traditional phishing campaigns and ML-enabled phishing campaigns. A meaningful difference would be one in which current cyber defense strategies would require significant improvements to mitigate the risk of ML-enabled phishing. Li and Kaloudi in their work, *The AI-Based Cyber Threat Landscape: A Survey,* make the following characterization of risks that impact the cybersecurity risk landscape as those that, "(i) expansion of existing threats, which deals with labor-intensive cyberattacks to achieve a large set of potential targets and low cost of attacks; (ii) introduction of new threats, which deals with tasks that would be impractical for humans; and (iii) change to the typical character of threats, which involves new attributes of finely targeted, automated, highly efficient, difficult to attribute, and large-scale attacks in the threat landscape" [29]. The risk of ML-enabled phishing is present, however, methods for mitigating that threat have been presented throughout our case studies. In DeepPhish, a publicly available dataset was used to determine URL filtering techniques and demonstrates that, just as adversaries can use models to extract features to bypass filtering, defenders can extract similar features to determine weak points for improvement [20]. In Turing in a Box, a method is presented for blue teams, building on the work in *Automatic Detection of Machine Generated Text: A Critical Survey* [18][25]. Given the significant concern around NLP for deception in phishing and other social engineering attacks [16], defenders will need to consider methods of identifying machine-generated text. Signatures can be developed for the identification of particular models, and some NLP makers may want to consider offering these identifiers as a means of mitigating the negative effects of their products.

## Synthesis Discussion

To understand the larger picture of ML-enabled phishing risk we also evaluated information across the different case studies and from other reviews/summary works. We ask what capabilities, intent, and access threat actors would need to exploit a vulnerability. We also ask what vulnerabilities are necessary for a well-positioned threat actor to take action. Additionally, we explore the consequences of these actions. Even if a well-positioned threat actor did exploit a vulnerability and used ML-enabled phishing to carry out an attack, it is important to understand if the defensive strategies needed to mitigate these attacks differ significantly from traditional phishing strategies. Finally, we briefly explore the consequences of this type of phishing. Below we have addressed several but not all of the questions posed. Some questions require further research outside the scope of this work, others are highly context-dependent. We believe additional research is needed to fully answer these questions in a meaningful way, but that these are the types of questions the security research community would benefit from answering to assess the overall risk of ML-enabled phishing.

Adapted from the AAAS-FBI-UNICRI National and Transnational Security Implications of Big Data in the Life Sciences Risk Assessment Framework

| Risk Assessment Framework for ML-enable Phishing | | |
|---|---|---|
| Probability of an Event Occuring | Characterization of the Adversary | · Which type(s) of adversary – lone actor, non-state/lone group, or nation-state – has the resources, technical expertise and skills, access to facilities and equipment, and motivation to attack with or exploit ML technologies for phishing?<br><br>· Have existing adversaries expressed interest in stealing, manipulating, or exploiting information using ML techniques?<br><br>· Does the adversary have the financial resources to carry out an attack or exploit systems using ML techniques?<br><br>· Does the adversary have access to a broader network of resources, including individuals, funding, equipment, etc?<br><br>· Does the adversary have the capabilities needed to access, manipulate and/or exploit ML technologies?<br><br>· Does the adversary have direct access to needed resources (data, ML development/tools, hardware)? If not, can the adversary gain access through hacking or elicitation?<br><br>· Would the adversary weigh the costs and benefits of achieving the attack? |

| | Characterization of the Adversary: Adversary Technical Expertise and Skills Needed | · What skills, expertise, and knowledge are needed to use data and ML techniques to enable a phishing campaign?<br><br>· What technologies and equipment are needed to use data and ML techniques to enable a phishing campaign?<br><br>· What skills, expertise, and knowledge are needed to exploit vulnerabilities in systems?<br><br>· What technologies, vulnerabilities, and other resources do the advisory need to leverage or access to carry out an attack?<br><br>· Is specialized access to technologies or systems needed to exploit some most, or all, vulnerabilities in the system? |
|---|---|---|
| | Vulnerabilities | · What technical measures exist to control access to and track users of ML technologies, data repositories, and hardware?<br><br>· What institutional measures exist to limit access to and identify users of the data systems and ML technologies?<br><br>· What are the current standards (technological and policy) used to prevent phishing, are they robust against ML advancements in phishing (if not why / in what way) |

| Consequences | Existing Counter-measures | · What countermeasures exist to address the vulnerabilities mentioned above?<br><br>· What countermeasures exist to prevent or mitigate the consequences of an attack?<br><br>· What measures exist to prevent or control the exploitation of information (data, open intel) and the malicious use of ML technologies? |
|---|---|---|
| | Outcomes of Successful Attack | · What are the possible immediate consequences of the attack to the economy, political systems, society, health, critical infrastructure, environment, and/or national security? **(hazard analysis)**<br>· What are the possible long-term consequences to the above?<br>· What is the scale of the immediate and longer-term consequences? |

## Characterization of the Adversary

Phishing, like many other attacker vs defender situations throughout cybersecurity, can be modeled as a two-player using game theory. We will refer to this as the 'phishing game'. In one of the case studies analyzed, *OFFENSIVE AI: UNIFICATION OF EMAIL GENERATION THROUGH GPT-2 MODEL WITH A GAME-THEORETIC APPROACH FOR SPEAR-PHISHING ATTACKS,* the authors present a game theory model for reasoning about adversary and defender strategies and possible action sets [30]. Phishing is a situation best modeled as a two-player game with incomplete information. In this simplistic model, adversaries have two main costs, the cost of preparing their attack and the cost if they are caught. Defenders also have two main costs, the cost of preparing for an attack or mitigating vulnerabilities, and the cost of a successful attack. Because of the asymmetric nature of information in this game, defenders are always at a disadvantage. Adversaries are always focused on remaining anonymous. Defenders, particularly in the case of organizations, are never anonymous. Individual targets, such as a single person targeted by a phishing scam, will have varying levels of anonymity depending on who they are. However, because the result of 'getting caught/identified' is not a cost for defenders, this variable does not matter at this level of analysis. It is a given that if a defender is being targeted by an adversary they have been identified. We have yet to find any literature that models the phishing game using varying levels of intel on a target as a variable of consideration.

In their work *A Game Theoretical Model for Anticipating Email Spear-Phishing Strategies,* the authors model the phishing game as a non-cooperative, non-zero sum, sequential game with incomplete and imperfect information [31]. This means defenders are assumed to not know historical actions taken by an attacker and are unaware of phishing techniques [31]. Additionally, this game is modeled sequentially, so defenders are only able to take an action once they receive communication, and both agents can be better or worse off in unequal proportion [31].

We use several models of the phishing game to better understand the factors that matter for adversaries and the strategies that best serve them. Using the results of these studies on phishing games, we can conclude the following about the priorities of and best strategies for adversaries [30][31].
- Adversaries can be modeled as instrumentally rational actors
- In repeated games, they are best served by sending an initial message how a low chance of detection relying on the principles of LSD, and the following up with a targeted attack
- While defense strategies are costly to defenders, adversaries who can remain undetected accrue little costs compared to the expected reward

The overwhelming majority of attackers are motivated by maximizing financial gains [19]. The demonstrations of ML-enabled phishing have come from the security research community who, as previously noted, have a different set of capabilities, access, and incentives. It is important to remember, however, that these approaches assume a rational attacker concerned only with maximizing utility. This assumes that, given the choice of exploiting a vulnerability using fewer resources than more, the attacker would choose to use fewer resources. However, not all actors may act rationally, and some attackers may miscalculate the amount of resources (time and tools) that are needed to maximize financial gains.

In an analysis of phishing adversaries, Halaseh and Alqatawna characterize the different threat actors based on nation and motivations. Their analysis shows the majority of phishing attacks are carried out for the purpose of committing cybercrime, followed by hacktivism, espionage, and cyber warfare [19]. Regardless of the end goal, they highlight that many well-positioned adversaries are able to work by themselves, in groups, with or without the support of a nation-state, and with techniques new and old [19]. Depending on the target(s) selected, all of these types of adversaries would theoretically then have the skills and resources required to leverage ML for their attacks. When it comes to using NLP it has never been easier, as the aptly named EasyNLP for example has made clear, for the average technologist to do so [32]. For those willing to pay, OpenAI offers several models with a quickstart tutorial that could let adversaries with a technical background start using NLP in an afternoon [33]. Even advanced features which allow users to do operations such as compare the similarity scores between two pieces of text are highly user friendly and accessible by design [33].

However, what would this process look like in practice? What would be involved if, for instance, an adversary wanted to impersonate a CEO of an organization in an email to a victim in the financial department of that organization by collecting the CEO's social media posts from Twitter to make a "sound-alike" message. First, they would have to bypass the site's anti-bot or anti-scraping security, which is achievable but a step in and of itself [34]. This established a process for adversaries of designing and executing some collection script, cleaning data, generating the scam, and editing and arranging the final text. If this process was done manually instead, and dozens of messages were collected and then fed into the OpenAI Davinci system by hand, they would then have a manual process of collecting, generating, editing, and arranging. It is difficult to imagine a situation where an ML-enabled process is significantly more efficient or advanced for this type of attack. There is no targeted phishing campaign without a process on the part of the adversary to identify and gather information on their target. An automated information retrieval system capable of collecting information on a target such as personality or tone of voice, beyond what is widely available as open-source tools, would most likely be a highly specialized tool used by well-financed agents. These types of tools may be used by an adversary for malicious purposes, but under what circumstances would this be both necessary and efficient? It has already been demonstrated that highly targeted techniques such as these are not necessary for high-profile targets. The scam that caused John Podesta, then manager of the Hillary Clinton presidential campaign, to reveal their login information had a scam that simply read, *Someone has your password* and included a shortened URL controlled by the Russian hacker group Fancy Bear [35].

In contrast, there are ways in which ML-enabled scamming can thwart existing policies and procedures meant to ensure security. In a recent vishing attack, adversaries used software to impersonate the CEO of a company and stole $243,000 [37]. Typically a phone call or other check of this sort is a standard method used to confirm important transfers at organizations. Any ML development that directly affects foundational policies and procedures (high likelihood, high impact) is a high-risk threat. This points again to the importance of having a systematic method of evaluating different ML capabilities against current cybersecurity practices in a timely fashion. Highly sophisticated voice mimicking software had been around for years prior to this attack.

Once a message successfully reaches its victim, the psychological manipulation that takes place is a matter of understanding the motivations of that person. Some scholars claim this is the advantage ML system will have over traditional methods. Tools used by professional recruiters in the field to psychologically profile targets for outreach, for example, were used in one case study. However, these types of tools have come under scrutiny and many have been shown to be biassed and unreliable [36]. It is difficult to determine if the average attacker will see a comparative advantage in profiling their targets using these tools given the additional work required to do so and their unreliability. In the vishing scam mentioned above, adversaries were able to manipulate a victim in real-time relying on the fundamental principles of creating a sense of urgency and impersonating someone with authority [37]. Although the tools are new, the basic elements motivating these scams will most likely remain the same.

## Existing Counter-measures

Just as adversaries can use large corpora of phishing emails to train models to execute attacks, defenders can use these corpora to train models that find patterns in phishing strategies [6]. In fact, the history of spam filters has shown the back and forth game adversaries and defenders must play when it comes to content. Spam is filtered by collecting a large body of spam and legitimate emails, and training an algorithm on individual words within these emails to determine which ones sound "spammy" [1][8]. New emails that arrive are given a probability measure of being spam and, given the mail user agent, may never even reach an inbox if they rise above a certain threshold [1]. Although those basic principles still apply, more advanced processes including email header detection, attachment vetting, and the relationship between words, not just of them, are now also employed [1]. As Rajivan and Gonzalez describe in their work *Creative Persuasion: A Study on Adversarial Behaviors and Strategies in Phishing Attacks,* as more detail on phishing strategies and their success rates emerges defenders will respond to them [26].

As discussed in the Research Questions Discussion section, it is possible for defenders to leverage methods of identifying machine-generated text to mitigate the risks associated with NLP [25]. Other countermeasures include using these tools that demonstrate these attacks for red team and pen testing operations, which is the intention behind the design and development of many of them before adversaries have a chance to use these techniques.

We can imagine a situation where, as defenders meet the current challenges of phishing, adversaries find it increasingly difficult to get victims to fall for their attacks. It is difficult to say in what ways adversaries would improve their tactics in response and if this improvement would involve machine learning. Our analysis points to the fact that ML will present different challenges for defenders overall, but is likely not the disruptive force some have suggested [2] [3] [4] [5] [18] [24]. Without a clear understanding of threat scenarios, we are sounding the alarm for a technology that may in fact not even be of interest to adversaries. Furthermore, although in no way conclusive, our research points to the fact that OSINT may be a significant risk to cyber defense operations overall.

## Future Work

There is not sufficient evidence to conclude that ML-enabled phishing campaigns are a high-priority threat for the defense community. We suggest more research is needed to draw meaningful conclusions about the risks this technology poses. More information looking in-depth at the strategies of current phishing campaigns and how ML might influence them is necessary. We suggest well-design studies with robust methodologies that would allow for testing individual aspects of a phishing campaign and how ML can enhance them.

## Works Cited

[1] M. Musser, and A. Garriott, "Machine Learning and Cybersecurity: Hype and Reality," Center for Security and Emerging Technology, June 2021. [Online]. doi: 10.51593/2020CA004.

[2] CISOMAG, "Artificial Intelligence as Security Solution and Weaponization by Hackers," *CISO MAG | Cyber Security Magazine,* December 9, 2019. [Online]. Available: https://cisomag.eccouncil.org/hackers-using-ai/.

[3] Darktrace, "The Next Paradigm Shift: AI-Driven Cyber-Attacks," *Darktrace Limited,* 2019. [Online]. Available: https://www.oixio.ee/sites/default/files/the_next_paradigm_shift_-_ai_driven_cyber_attacks.pdf.

[4] D. Drinkwater, "6 Ways Hackers Will Use Machine Learning to Launch Attacks," CSO Online, January 22, 2018. [Online]. Available: https://www.csoonline.com/article/3250144/6-ways-hackers-will-use-machine-learning-to-launch-attacks.html.

[5] C. Beek, et al., "McAfee Labs 2017 Threats Predictions Report," *McAfee Labs,* November 2016. [Online]. Available: https://www.mcafee.com/enterprise/en-us/assets/reports/rp-threats-predictions-2017.pdf.

[6] Federal Bureau of Investigation Internet Crime Complaint Center, "2021 IC3 Annual Report," 2021. [Online]. Available: https://www.ic3.gov/Media/PDF/AnnualReport/2021_IC3Report.pdf.

[7] E. EH. Lastdrager, "Achieving a consensual definition of phishing based on a systematic review of the literature," *Crime Science* 3, no. 1:1-10, 2014. [Online]. Available: https://crimesciencejournal.biomedcentral.com/articles/10.1186/s40163-014-0009-y.

[8] F. Bruton, *Spam: A Shadow History of the Internet.* The MIT Press. 2013. [Online]. Available: https://doi.org/10.7551/mitpress/9384.001.0001.

[9] V. K. Devendran, H. Shahriar, and V. Clincy, "A Comparative Study of Email Forensic Tools," *Journal of Information Security,* vol. 06 (02): 111–17, 2015. [Online]. Available: https://doi.org/10.4236/jis.2015.62012.

[10] E. O. Yeboah-Boateng and P. M. Amanor, "Phishing, SMiShing & Vishing: An Assessment of Threats against Mobile Devices," *Journal of Emerging Trends in Computing and Information Sciences,* vol. 5, no. 4, p. 11, 2014. [Online]. Available: https://e-tarjome.com/storage/btn_uploaded/2020-09-12/1599891065_11216-etarjome%20English.pdf.

[11] Z. Alkhalil, C. Hewage, L. Nawaf, and I. Khan, "Phishing Attacks: A Recent Comprehensive Study and a New Anatomy," *Frontiers in Computer Science,* 3. 2021. [Online]. Available: https://www.frontiersin.org/article/10.3389/fcomp.2021.563060.

[12] T. Stojnic, D. Vatsalan, and N. A. G. Arachchilage, "Phishing email strategies: Understanding cybercriminals' strategies of crafting phishing emails," *SECURITY AND PRIVACY*, vol. 4, no. 5, p. e165, 2021, [Online]. doi: [10.1002/spy2.165](10.1002/spy2.165).

[13] W. Pearce, N. Landers. "1 09 42 The answer to life the universe and everything offensive security Will Pearce Nick Landers." Derbycon. 2019. [Online]. Available: https://www.youtube.com/watch?v=CsvkYoxtexQ

[14] A. Ferreira, L. Coventry, and G. Lenzini, "Principles of Persuasion in Social Engineering and Their Use in Phishing," in *Human Aspects of Information Security, Privacy, and Trust*, Cham, 2015, pp. 36–47. [Online]. doi: [10.1007/978-3-319-20376-8_4](10.1007/978-3-319-20376-8_4).

[15] C. Hadnagy, and M. Fincher. *Phishing Dark Waters: The offensive and defensive sides of malicious emails.* John Wiley & Sons, 2015.

[16] W. D. Heaven, "OpenAI's new language generator GPT-3 is shockingly good—and completely mindless," MIT Technology Review, July 20, 2020. [Online]. Available: https://www.technologyreview.com/2020/07/20/1005454/openai-machine-learning-language-generator-gpt-3-nlp/.

[17] W. Knight, "GPT-3 Can Write Disinformation Now—and Dupe Human Readers," WIRED, May 24, 2021. [Online]. Available: https://www.wired.com/story/ai-write-disinformation-dupe-human-readers/ .

[18] E. Lim, G. Tan, T. K. Hock, T. Lee,"Turing in a Box: Applying Artificial Intelligence as a Service to Targeted Phishing and Defending against AI-generated Attacks," GovTech Singapore, 2021. [Online]. Available: https://i.blackhat.com/USA21/Wednesday-Handouts/US-21-Lim-Turing-in-a-Box-wp.pdf.

[19] R. A. Halaseh and J. Alqatawna, "Analyzing CyberCrimes Strategies: The Case of Phishing Attack," in *2016 Cybersecurity and Cyberforensics Conference (CCC)*, Aug. 2016, pp. 82–88. doi: [10.1109/CCC.2016.25](10.1109/CCC.2016.25).

[20] A. C. Bahnsen, I. Torroledo, L. D. Camacho, and S. Villegas, "DeepPhish: Simulating Malicious AI," *2018 APWG symposium on electronic crime research (eCrime)*, May 201, .pp. 1-8. [Online]. Available: https://albahnsen.files.wordpress.com/2018/05/deepphish-simulating-malicious-ai_submitted.pdf.

[21] A. P. Siddaway, A. M. Wood, and L. V. Hedges, "How to Do a Systematic Review: A Best Practice Guide for Conducting and Reporting Narrative Reviews, Meta-Analyses, and Meta-Syntheses," *Annu. Rev. Psychol.*, vol. 70, no. 1, pp. 747–770, Jan. 2019, doi: [10.1146/annurev-psych-010418-102803](10.1146/annurev-psych-010418-102803).

[22] E. M. Beller *et al.*, "PRISMA for Abstracts: Reporting Systematic Reviews in Journal and Conference Abstracts," *PLOS Medicine*, vol. 10, no. 4, p. e1001419, Apr. 2013, doi: [10.1371/journal.pmed.1001419](10.1371/journal.pmed.1001419).
25

[23] American Association for the Advancement of Science, "National and Transnational Security Implications of Big Data in the Life Sciences, " AAAS-FBI-UNICRI, p. 47, 2014. [Online]. Available: https://www.aaas.org/sites/default/files/AAAS-FBI-UNICRI_Big_Data_Report_111014.pdf.

[24] J. Seymour and P. Tully, "Generative Models for Spear Phishing Posts on Social Media," *arXiv:1802.05196 [cs, stat]*, Feb. 2018, [Online]. Available: http://arxiv.org/abs/1802.05196.

[25] G. Jawahar, M. Abdul-Mageed, and L. V. S. Lakshmanan, "Automatic Detection of Machine Generated Text: A Critical Survey," *arXiv:2011.01314 [cs]*, Nov. 2020, [Online]. Available: http://arxiv.org/abs/2011.01314

[26] P. Rajivan and C. Gonzalez, "Creative Persuasion: A Study on Adversarial Behaviors and Strategies in Phishing Attacks," *Frontiers in Psychology*, vol. 9, 2018, [Online]. Available: https://www.frontiersin.org/article/10.3389/fpsyg.2018.00135

[27] S. C. Matz, R. E. Appel, and M. Kosinski, "Privacy in the age of psychological targeting," *Current Opinion in Psychology*, vol. 31, pp. 116–121, 2020, doi: https://doi.org/10.1016/j.copsyc.2019.08.010.

[28] M. Funk, "Opinion | Cambridge Analytica and the Secret Agenda of a Facebook Quiz," *The New York Times*, Nov. 19, 2016. [Online]. Available: https://www.nytimes.com/2016/11/20/opinion/cambridge-analytica-facebook-quiz.html

[29] N. Kaloudi and J. Li, "The AI-Based Cyber Threat Landscape: A Survey," *ACM Comput. Surv.*, vol. 53, no. 1, pp. 1–34, Jan. 2021, doi: 10.1145/3372823.

[30] H. Khan, M. Alam, S. Al-Kuwari, and Y. Faheem, "OFFENSIVE AI: UNIFICATION OF EMAIL GENERATION THROUGH GPT-2 MODEL WITH A GAME-THEORETIC APPROACH FOR SPEAR-PHISHING ATTACKS," in *Competitive Advantage in the Digital Economy (CADE 2021)*, Jun. 2021, vol. 2021, pp. 178–184. doi: 10.1049/icp.2021.2422.

[31] F. Tchakounte, V. Nyassi, D. Danga, K. Udagepola, and M. Atemkeng, "A Game Theoretical Model for Anticipating Email Spear-Phishing Strategies," *ICST Transactions on Scalable Information Systems*, p. 166354, Jul. 2018, doi: 10.4108/eai.26-5-2020.166354.

[32] "EasyNLP: An Easy-to-use NLP Toolkit," *Python Awesome*, Apr. 08, 2022. [Online]. Available: https://pythonawesome.com/easynlp-an-easy-to-use-nlp-toolkit/.

[33] "OpenAI API." https://beta.openai.com

[34] B. Amin Azad, O. Starov, P. Laperdrix, and N. Nikiforakis, "Web Runner 2049: Evaluating Third-Party Anti-bot Services," in *Detection of Intrusions and Malware, and Vulnerability Assessment*, Cham, 2020, pp. 135–159. doi: 10.1007/978-3-030-52683-2_7.

[35] "The phishing email that hacked the account of John Podesta," CBS News, [Online]. Available: https://www.cbsnews.com/news/the-phishing-email-that-hacked-the-account-of-john-podesta.

[36] M. R. and S. Barocas, "Challenges for mitigating bias in algorithmic hiring," *Brookings*, Dec. 06, 2019. [Online]. Available:

https://www.brookings.edu/research/challenges-for-mitigating-bias-in-algorithmic-hiring/

[37] C. Stupp, "Fraudsters Used AI to Mimic CEO's Voice in Unusual Cybercrime Case," *Wall Street Journal*, Aug. 30, 2019. [Online]. Available:

https://www.wsj.com/articles/fraudsters-use-ai-to-mimic-ceos-voice-in-unusual-cybercrime-case-11567157402